\documentclass[prd,superscriptaddress,amsfonts,amssymb,
amsmath,showkeys,showpacs,twocolumn,epsfig,graphics,floatfix]{revtex4}
\usepackage{amsmath}
\usepackage{epsfig,graphics}
\date{\today}
\usepackage{graphics}

\begin{document}

\author{George E.\ A.\ Matsas}
\email{matsas@ift.unesp.br}
\author{Vicente Pleitez}
\email{vicente@ift.unesp.br}
\affiliation{Instituto de F\'\i sica Te\'orica, Universidade
Estadual Paulista,
R.\ Pamplona 145, 01405-900 - S\~ao Paulo, SP, Brazil}
\author{Alberto Saa\footnote{On leave of absence from UNICAMP, Campinas, SP, Brazil.}}
\email{asaa@ime.unicamp.br}
\affiliation{Centro de Matem\'atica, Computa\c c\~ao e Cogni\c c\~ao,
Universidade Federal do ABC, 09210-170, Santo Andr\'e, SP, Brazil}
\author{Daniel A.\ T.\ Vanzella}
\email{vanzella@ifsc.usp.br}
\affiliation{Instituto de F\'\i sica de S\~ao Carlos, Universidade de S\~ao Paulo,\\
Av. Trabalhador S\~ao-carlense, 400, C.\ P.\ 369, 13560-970,
S\~ao Carlos, SP, Brazil}

\title{The number of dimensional fundamental constants}


\begin{abstract}
We revisit, qualify, and {\em objectively} resolve the seemingly controversial question
about what is the number of {\em dimensional} fundamental constants in Nature. For
this purpose, we only assume that all we can directly measure are space and time
intervals, and that this is enough to evaluate any physical observable. We conclude
that the {\em number of dimensional fundamental constants is two}. We emphasize
that this is an {\em objective result} rather than a ``philosophical opinion",
and we let it clear how it could be refuted in order to prove us wrong.
Our conclusion coincides with Veneziano's string-theoretical one but our arguments
are not based on any  particular theory. As a result, this implies that one of
the three usually considered fundamental constants $G$, $c$ or $h$ can be eliminated
and we show explicitly how this can be accomplished.
\end{abstract}

\maketitle

\section{Introduction}
\label{sec:intro}

The problem of how many dimensional fundamental constants are in Nature
has been subject of debate for a long time (see, e.g.,
\cite{Levy77}-\cite{Wilczek} and references therein). In the beginning
of the XX century, Planck stated that four dimensional constants were
necessary and sufficient to describe all physical phenomena~\cite{Planck}.
He named these constants $a$, $b$, $c$, and $f$, which turned out to be
related to the modern constants $h$, $k$ and $G$ by $a=h/k$, $b=h$ and
$f=G$. The symbol $c$ has been kept since then to represent the speed of light.
Planck showed, furthermore, that these constants could be combined in
order to get a ``natural" system of three basic units: units of
length $\ell_{P} =  \sqrt{G\hbar/c^3}$,
time $t_{P} = \sqrt{G\hbar/c^5}$,
and mass $m_{ P} = \sqrt{\hbar c/G}$.
(Incidentally, their numerical values were close to the  set of units
previously suggested by Stoney in the context of electromagnetism~\cite{Stoney}
even before the introduction of the Planck constant $h$.)

A closer analysis of the four Planck's original dimensional constants
has revealed that the combination $b/a$, namely $k$, turns out to be
a conversion factor between temperature and energy units, leading to
the well accepted conclusion that the Boltzmann constant has its
origin in the historical fact that temperature was not early recognized
as a manifestation of kinetic energy. Had temperature been defined, at
early times, as twice the mean energy stored in each degree of freedom
of a system in thermal equilibrium (whenever the equipartition energy
principle is valid) and the Boltzmann constant would have the
dimensionless value 1 (i.e., it would not have been introduced at all),
in which case entropy would be also dimensionless.

The view that $G$, $c$ and  $h$ would be the three fundamental constants of
Nature is often expressed by means of the so called ``cube of natural
units" or ``cube of theories" first introduced by Gamov, Ivanenko and
Landau in the late twenties (see, e.g., \cite{Okun91} and references therein),
where the three dimensional constants
involved in the construction of the Planck basic units appear explicitly. The
vertices of the cube would correspond to certain limiting regimes of the physical
laws. For instance, the origin $(0,0,0)$ would correspond to non-relativistic
mechanics, $(c^{-1},0,0)$ to Special Relativity, $(0,h,0)$ to non-relativistic
quantum mechanics, $(c^{-1},0,G)$ to General Relativity and so on~\cite{Okun91}
(see left-hand side of Fig.~\ref{squashing}).

Such a conception has been recently challenged. Veneziano has
concluded through string-theoretical arguments that the number of dimensional
fundamental constants would be two, while Duff has advocated for none at
all~\cite{Duffetal02}. This is quite unacceptable that this question is still
controversial. The problem here does not concern the fact of not having
an answer to some {\em objective} question but having different answers to the same
question. Here we make a clear
statement of the problem and present what we believe to be a final
solution: {\em the number of fundamental dimensional constants is two}.
We emphasize that our conclusion can be objectively refuted if it is shown
that there exists some observable measured in laboratory that cannot be
expressed  in a basis of {\em two} independent  observables as shown in
Eq.~(\ref{o1o2}). What we present here is, thus, an objective conclusion
rather than a philosophical opinion.

\section{Space, time, and nothing else}
\label{sec:ST}

Let us begin by clearly stating the points upon which our line of reasoning is
based:
\begin{itemize}

\item[{\bf 1.}] Without going into the subtleties of precisely defining what a
physical theory {\it is}, we only assume that any ``good'' physical theory should
fulfill the following minimum criteria:
(i) list the observables it deals with,
(ii) prescribe how to measure these observables, and
(iii) provide self-consistent relations among them (the ``physical laws'');

\item[{\bf 2.}] These relations are then tested against experimental data
obtained through properly chosen processes, which we {\it assume} to take
place in the spacetime; {\it eventually, all we can directly measure are
space and time intervals}. In particular, this {\it implies} that one only
needs two units to express all measurements.
\end{itemize}

The {\it basic units} of space and time, say $\sigma$ and $\tau$,
respectively, can be chosen in a quite arbitrary way, and once this
is done, any observable ${\cal O}_i$ can be expressed as
\begin{eqnarray}
{\cal O}_i=\Omega_{i}\,\tau^{{\alpha}_i}\sigma^{{\beta}_i},
\label{obs1}
\end{eqnarray}
with $\Omega_i$, $\alpha_i$, and $\beta_i$ being real (dimensionless)
numbers, and $i$ belonging to some index set $\cal I$. For the sake of
notation simplicity, the index $i$ gives information not only about the
physical quantity being considered (energy, spin, \ldots) but also about
the state of the system (no matter how the theory chooses to describe it).
As a result, $\Omega_i$, for given $i$, is indeed a real number
instead of a real-valued function.

It is in order now to stress that one can select any pair $o_1$ and $o_2$
of {\it independent} observables to express all the other ones as follows.
Since $o_1, o_2 \in \{ {\cal O}_i\}_{i\in {\cal I}}$, let us cast them as
\begin{eqnarray}
o_1&=&\Omega_{o_1}\tau^{a_1}\sigma^{b_1}\; ,
\label{o1}
\\
o_2&=&\Omega_{o_2}\tau^{a_2}\sigma^{b_2}\;,
\label{o2}
\end{eqnarray}
where by independent we mean that $(a_1,b_1)$ and $(a_2,b_2)$ form a basis of
${\mathbb R}^2$. Then, we can solve Eqs.~(\ref{o1}) and (\ref{o2}) for
$\tau$ and $\sigma$ and rewrite Eq.~(\ref{obs1}) associated with all
other observables as
\begin{eqnarray}
{\cal O}_i={\widetilde \Omega}_i o_1^{\mu_i}o_2^{\nu_i}\; ,
\label{o1o2}
\end{eqnarray}
where ${\widetilde \Omega}_i$, $\mu_i$ and $\nu_i$ are clearly real numbers, namely,
\begin{eqnarray}
{\widetilde \Omega}_i
& = &  \Omega_i \;
  \Omega_{o_1}^{(-b_2 \alpha_i + a_2 \beta_i)/W} \;
  \Omega_{o_2}^{( b_1 \alpha_i - a_1 \beta_i)/W}\; ,
\nonumber\\
\mu_i & = & (b_2 \alpha_i - a_2 \beta_i)/W \; ,
\nonumber \\
\nu_i & = & (-b_1 \alpha_i + a_1 \beta_i)/W
\nonumber
\end{eqnarray}
and $W \equiv a_1 b_2 - b_1 a_2 $. The fact that
$W\neq 0$ is guaranteed by our requirement
that $o_1$ and $o_2$ be independent.
Although the choice $\{o_1,o_2\}\subset \{{\cal O}_i\}_{i\in {\cal I}}$
is arbitrary, there may be more convenient ways (possibly theory-dependent)
of selecting this basic set. Indeed, the
speed of light $c$ and the transition time $t_{Cs}$ between certain energy
states of the Cesium atom, which are presently adopted to define the standard
units of space and time~\cite{Mohretal05}, could be naturally chosen to
constitute the basic set.

At this point, let us clearly distinguish what we are saying from what we are
{\it not} saying. Our conclusion that all the observables ${\cal O}_i$
can be expressed in terms of only two basic dimensional observables does {\it not}
imply the existence of some ``final'' theory able to {\em predict}
all dimensionless real numbers ${\widetilde \Omega}_i$ appearing in Eq.~(\ref{o1o2})
(no matter how desirable it may be).
For instance, in the Standard Model the total number of dimensional and
dimensionless parameters which should be provided as input in
order to predict all the other observables is much larger than two. However,
one can choose any pair of independent dimensional constants from this set and
express all the other ones (and, consequently, all observables) in terms of this
pair.

\section{Measuring mass with clocks and rulers}
\label{sec:mass}

A question which can be raised is how observables which are usually
written not only in terms of space and time units, e.g.\ mass, fit
into this scheme. We exhibit next two protocols where the mass unit $M$ (g,
pounds, \ldots) is solely expressed in terms of units of space and
time. For the sake of simplicity, in both protocols we start
assuming the CGS system, where all observable quantities ${\cal D}_i$ are
expressed in terms of units of space $L$~(cm), time $T$~(s), and mass $M$~(g):
\begin{eqnarray}
{\cal D}_i=\Delta_{i} \; T^{{\alpha}_i}L^{{\beta}_i}M^{{\gamma}_i}\;,
\label{calD}
\end{eqnarray}
where $\Delta_i$, $\alpha_i$, $\beta_i$, and $\gamma_i$ are real numbers.

\begin{itemize}
\item \emph{$G$ protocol}:
Multiply Eq.~(\ref{calD}) by $G^{\gamma_i}$, where $G$ is the Newton's
constant, and identify ${\cal D}_i G^{\gamma_i}$ as the observable
${\cal O}_i^{(G)}$ which appears in Eq.~(\ref{obs1}) (and fulfills our
conditions {\bf 1.}--{\bf 2.}). [Here the superscript $(G)$ is introduced
only to indicate the protocol used to cast the observable in the
form given by Eq.~(\ref{obs1}).]
By using the $G$ protocol in all observables involving the $M$ unit
and rewriting the physical laws in terms
of ${\cal O}_i^{(G)}$ rather than ${\cal D}_i$, we end up (i)~vanishing
the unit $M$ (g) from the observables, (ii)~vanishing the constant $G$
from all physical laws, and (iii)~with masses being measured in units
of ${\rm cm}^3/{\rm s}^2$. For instance, by applying the $G$ protocol to
the original Newton's gravitational law $g=-Gm/d^2$, we get
\begin{eqnarray}
g=-\, {m^{(G)}}/{d^2}\;,
\label{FG}
\end{eqnarray}
where the units of $m^{(G)}\equiv m \, G$ is $L^3/T^2$.  In this sense,
$G$ plays the role of a conversion factor (from ${\rm cm}^3/{\rm s}^2$ to g)
as much as does the Boltzmann constant $k$ (from erg to Kelvin). Indeed,
$G$ could have never been introduced once the mass unit were properly
defined (see Sec.~11 of Ref.~\cite{Buckingham14} and Ref.~\cite{Okunetal02}).
We see from Eq.~(\ref{FG}) that this procedure of
eliminating the unit of mass $M$ leads $m^{(G)}$ to be directly interpreted
as (and not only related to) {\it active gravitational} mass,
and that it can be obtained simply through space and time measurements
by determining the acceleration $g$ induced on test particles lying at
a distance $d$.
Moreover, we note for further purposes that the $G$ protocol applied to
(the observable) Planck's constant $h$ leads to
\begin{eqnarray}
h^{(G)}\equiv h\,G\;,
\end{eqnarray}
whose dimension is $L^5/T^3$.

\item \emph{$h$ protocol}:
Divide Eq.~(\ref{calD}) by $h^{\gamma_i}$, where $h$ is Planck's constant,
and identify ${\cal D}_i/h^{\gamma_i}$ as the observable ${\cal O}_i^{(h)}$
which satisfies Eq.~(\ref{obs1}). By rewriting the physical laws in terms of
${\cal O}_i^{(h)}$ rather than ${\cal D}_i$, we end up (i)~vanishing
the unit $M$ (g) from the observables as before, (ii)~vanishing the
constant $h$ from all physical laws, and (iii)~with masses being measured
in units of ${\rm s}/{\rm cm}^2$. For instance, by applying the $h$ protocol
to the original Compton scattering formula $\Delta \lambda=[h/(m c)]
(1-\cos\theta)$, we get
\begin{eqnarray}
\Delta \lambda =[m^{(h)}c]^{-1}(1-\cos\theta)\;,
\label{Eh}
\end{eqnarray}
where the unit of $m^{(h)}\equiv m/h$ is $T/L^2$.  In this sense, $h$ plays
the role of a conversion factor (from ${\rm s}/{\rm cm}^2$ to g). In this
protocol the inverse of the Compton length can be seen to be
naturally associated with {\it inertia} of elementary particles through
the Compton effect. Here $m^{(h)} c$ (and therefore $m^{(h)}$)
can be directly measured using clocks and rulers by determining the wavelength
change $\Delta \lambda$ of a photon scattered by an angle $\theta$.
We note that the $h$ protocol applied to Newton's constant $G$ leads to
\begin{eqnarray}
G^{(h)}\equiv G h \; ;
\end{eqnarray}
Note that $G^{(h)}= h^{(G)}$. (A sort of implementation of the $h$ protocol
can be found in Refs.~\cite{Hoyle}-\cite{Wignall}.)
\end{itemize}
Some physical observables of interest in the different protocols are
shown in Table~\ref{PQ}.
\begin{table}[h]
\begin{tabular}{|c|c|c|}
\hline
  & $G$ protocol &  $h$ protocol      \\
\hline
$G$   &   $G^{(G)} =  1$
      &   $G^{(h)} =  4.42 \times 10^{-34} \; {\rm cm}^5/{\rm s}^3$\\
\hline
$h$   &   $h^{(G)} =  4.42 \times 10^{-34} \; {\rm cm}^5/{\rm s}^3$
      &   $h^{(h)} =  1$\\
\hline
$m_e$ & $m_e^{(G)} =  6.08 \times 10^{-35} \;{\rm cm}^3/{\rm s}^2$
      & $m_e^{(h)} =  1.38 \times 10^{  -1} \;{\rm s}/{\rm cm}^2$\\
\hline
$ e $ &   $e^{(G)} =  1.24 \times 10^{-13} \;{\rm cm}^3/{\rm s}^2$
      &   $e^{(h)} =  0.59 \times 10^{  4} \;{\rm cm}^{1/2}/{\rm s}^{1/2}$\\
\hline
\end{tabular}
\caption{The values of $G$, $h$, and the electron mass and charge are presented in
the $G$~and $h$~protocols. It is interesting to note that
$h^{(G)}=G^{(h)}$.}
\label{PQ}
\end{table}

Rewriting the physical laws in terms of Eq.~(\ref{o1o2})
rather than Eq.~(\ref{calD}) does not change any of their predictions.
Nevertheless, it can shed new light on some conceptual issues. Next, we discuss
and resolve in this context the much-debated question about what is the
number of dimensional fundamental constants in Nature.

\section{Physical Insights}
\label{sec:insights}

In the previous section we presented
two protocols in which mass can be determined through
space and time measurements alone. In order to do so,
each one of those protocols made use of a
specific law relating mass with space and time intervals. However,
since we claim that our main conclusion (regarding the number
of dimensional fundamental constants) is general and
theory-independent, one might wonder
``what if Nature did not comply with any such a law?'' For the sake of concreteness,
let us imagine that Newtonian mechanics in the {\it absence} of gravity
were all that there was to the laws of Nature. How would
our arguments apply in this case?

Firstly, it is important
(though obvious) to point out that Nature would look completely different.
In that case, (inertial)
mass would no longer be determined in terms of space and time measurements
and a {\it standard} of mass, let us say the {\it kilogram}, might be introduced.
Any mass would then be determined as some multiple of this standard mass through,
e.g., colision experiments (recall that gravity is not available). Our point,
however, is that in this case mass is no longer an {\it observable} as defined in
Eq.~(\ref{obs1}), and, therefore, the laws of Nature can be rewritten in
such a way to completely avoid the appearance of mass. This is certainly possible
according to our
definition of ``physical law'' given in point {\bf 1.} of Sec.~\ref{sec:ST},
although they may look more complicated and appear in a larger number
in this rewritten form. Once that is done, we are left again with only space
and time units.

For those (like some of the authors)
who would rather take the philosophical stand that even in the context above
(Newtonian mechanics with {\it no} gravity) mass should be considered
as ``observable'' (in some extended sense) due to the simplification it
brings to the form of the physical laws, we point out that the need for an independent
standard in this case reflects the fact that mass would be determined only up to a multiplicative
constant: no mass scale would be privileged (Nature would be completely
different indeed).
Therefore, this new standard (introduced only to comply with someone's prejudice)
would not appear in any ``objective fundamental constant'' (i.e., those
which determine
``fundamental scales'', as most people seem to interpret what a
fundamental constant is). Interestingly enough,
in this same context a similar argument can be applied to space and
time units to show that Newtonian mechanics in the absence of gravity
has no ``objective fundamental constants'' at all (no fundamental space or time
scales). The fact that we experience space and time as distinct entities and
that Nature does not present itself invariant under rescale of these quantities
seem to indicate that there should indeed exist fundamental scales of space and time.
It is a remarkable fact that we have already reached an understanding of
the physical laws from which these scales can be read out.

Since Nature does provide ways of measuring mass
in terms of space and time intervals (the
$G$ and $h$~protocols), the main relevance of the discussion above
lies in the fact that it can be applied to any ``observable'' (in the extended sense)
which may appear in some (still unknown) physical law: the ``observable'' is
either determined in terms of space and time measurements (a true observable)
or it cannot appear in determining fundamental constants which set scales
in Nature.

Up to this point, our discussion was quite general and our conclusions
followed directly from assumptions {\bf 1.-2.} of Sec.~\ref{sec:ST}.
Here, we raise accordingly some speculative physical considerations, which
may be of some relevance. For this purpose, we choose the protocol $G$.

Firstly, we note that the protocol $G$ frees classical mechanics from its
only (so called) ``fundamental parameter", namely, Newton's gravitational
constant. This is very convenient since the value of $G$ (like the value
of $k$) does not determine any scale for new physics, in contrast to,
e.g., $c$ and $h$, which fix a velocity and an angular momentum scale,
respectively, where relativistic and quantum mechanical effects become
important. In the $G$ protocol  it becomes clear that classical
mechanics does not have any prefered basis $\{ o_1, o_2 \}$ of independent
observables (see the origin of the plane of theories in Fig.~\ref{squashing}).
Newton's gravitational and second laws are written in this context as
\begin{equation}
F_{g}^{(G)} = m^{(G)} M^{(G)} /r^2,\;\;\;{\rm and}\;\;\; F^{(G)}_{I} = m^{(G)}a,
\label{Newton}
\end{equation}
respectively. By comparing Eq.~(\ref{Newton}) with the Coulomb law
$
F_{C}^{(G)} = q^{(G)} Q^{(G)} /r^2
$ we see
that (gravitational) mass and electric charge have the same
status as coupling constants of
the respective interactions, as they should. The fact that
gravitational mass happens to give also a measure of {\it inertia}
is a separate, more profound issue, only partially addressed by General
Relativity~\footnote{The equality of inertial and {\it passive} gravitational
mass finds an elegant explanation in the context of
(and actually led to) metric theories of gravity.
The relation between inertial and {\it active} gravitational mass,
however,
is more subtle and, to some extent, still
unaddressed. In modern terms,
{\it why is the energy content of a system so intimately tied to its potential
to curve the background spacetime?} Answering this question is equivalent to
{\it explaining} Einstein equations, which possibly requires (or
would lead to)
a ``quantum gravity'' theory.}.

The actual bias driven by our present theories suggests that
a natural basis of independent observables would be $\{ c, h^{(G)} \}$.
The vertices $(c^{-1}=0, h^{(G)})$ and $(c^{-1}, h^{(G)}=0)$
in the plane of theories (see Fig.~\ref{squashing}) represent non-relativistic
and classical physics, respectively. It is interesting to note that in this
scenario what we usually denominate (a) {\em quantum gravity} (whatever
it is) and (b) {\em quantum  field theory} would dwell at the same vertex
$(c^{-1}, h^{(G)} )$ of Fig.~\ref{squashing}. This would be reflecting
the expected feature that a complete theory (which we call here
{\em quantum relativity}) describing consistently all interactions
(including gravity) should contain (a) and (b). We note that
quantum relativity effects can appear at quite low energies.
The Hawking temperature
$$
(kT_H)^{(G)} = \frac{\hbar^{(G)} c^3}{8 \pi M^{(G)}}
$$
associated with the evaporation of (static) black holes with mass $M^{(G)}$
is a good example of it. If $c$
were substantially larger, $(kT_H)^{(G)}$  would be able to be
trivially tested in stellar-size black holes.
A distinct point that becomes obvious in the context of
the $G$ protocol is that an electron has much more electric than
``gravitational" charge: $e^{(G)}= 2 \times 10^{21} m_e^{(G)}$. On the other
hand General Relativity tells us that if a (classical) black hole is
given more charge than mass it becomes a naked singularity.
Naked singularities are expected to be understood only in the context of
quantum gravity. If quantum gravity and quantum field theory
are both different aspects of a single quantum relativity theory, it is
possible that naked singularities and elementary particles be understood
by the same token. This is not odd if one notices that the Planck scale
for the angular momentum is  precisely given by $h^{(G)}$.

\section{Conclusions}
\label{sec:discussion}

As a result of our conclusion that all observables can be expressed
as in Eq.~(\ref{o1o2}), {\em any physical theory fulfilling
conditions} {\bf 1.}--{\bf 2.}\ {\em will not require more than
a pair of independent dimensional constants to be expressed}.
This pair of observables is what one usually denominates {\it fundamental
(dimensional) constants}. Both protocols
that we have presented favor $c$ and $h^{(G)}$ ($= G^{(h)}$)
to constitute this pair but other protocols could be easily devised
(e.g., replacing $h$ by $e^2$ in the $h$ protocol). From a theory-independent
perspective, no protocol can be privileged. However, some distinction
can be made if we allow for biases coming from the present theories.
In this sense, both the $G$~and $h$~protocols exhibit the nice feature
of selecting as fundamental constants $c$ and $h^{(G)}$, which define
the regimes where relativistic and quantum effects become important,
respectively. Our conclusion that {\em two} ($c$, $h^{(G)}$) rather than
{\em three} ($c$, $h$, $G$) dimensional fundamental constants suffice can
be nicely represented as  in Fig.~\ref{squashing} by the squashing of
the cube of theories. Clearly, other choices of fundamental constants
may become more convenient depending on future developments of our theories.
Eventually, for a full quantum gravity (or ``quantum relativity'') theory,
the following pair may be more convenient: $t_{P}= (h^{(G)}/c^5)^{1/2}$ and
$\ell_{P}= (h^{(G)}/c^3)^{1/2}$ (see Fig.~\ref{TPLP}).

In order to avoid any misunderstanding, let us clarify one point
which might cause some confusion. After adopting the
$G$~(or $h$)~protocol to eliminate the mass unit $M$ from
all quantities in Eq.~(\ref{calD}), one could easily envisage
an extra protocol  to carry out the elimination of one
of the remaining units $T$ {\it or} $L$, e.g., a ``$c$~protocol'',
in which case all quantities would be expressed accordingly
as multiples of some power of $L$ {\it or} $T$ alone. In fact, one
could proceed even further and use the other fundamental constant
$h^{(G)}$ (after properly redefined through the hypothetical
$c$~protocol) and vanish the sole unit left. In this case, all
quantities would be dimensionless numbers. Although this is true,
it does {\it not} imply that the number of dimensional fundamental
constants would be one or zero~\cite{explanation}. We must recall that
our line of reasoning is based on points {\bf 1.--2.}\ of Sec.~\ref{sec:ST},
and {\it not} on the existence of some protocol to lower the number
of units in physical equations. The latter only eliminates
unnecessary structures once the number of dimensional fundamental
constants is established. After all, our conclusion that the
number of dimensional fundamental constants is two
comes from the following facts: (i)~space and time are distinct
entities (a fact which should be accounted for in any ``complete
theory'') and (ii)~combined measurements of space and
time suffice to characterize any physical system. Points (i)~and
(ii)~above set the
{\em lower} and {\em upper} bounds, respectively, for the number
of dimensional fundamental constants to be {\em two}.
Our conclusions are in agreement with Veneziano (see
Ref.~\cite{Duffetal02}) but our arguments are model independent.

We emphasize once again that our conclusion above can be objectively
refuted if it is shown that there exists some observable measured in
laboratory that cannot be expressed  in a basis of {\em two} independent
observables as shown in Eq.~(\ref{o1o2}). This is, thus, an objective
conclusion rather than a philosophical opinion. A number of valuable
papers reformulating physical theories in a more transparent or
elegant way can be found in the literature. We hope that our paper is
appreciated in the same lines with the difference that it deals with
the whole set of physical theories.

\acknowledgments

The authors would like to thank O.\ Saa for calling their
attention to Ref.~\cite{Buckingham14} and discussions. We are
also grateful to E.\ Abdalla, L.\ C.\ B.\ Crispino, C.\ O.\ Escobar,
C.\ A.\ S.\ Maia and J.\ C.\ Montero for conversations. The authors
are also grateful to CNPq, FAPESP, and UFABC for financial support.

\newpage
\begin{widetext}
\begin{figure}[h]
\mbox{\epsfig{file=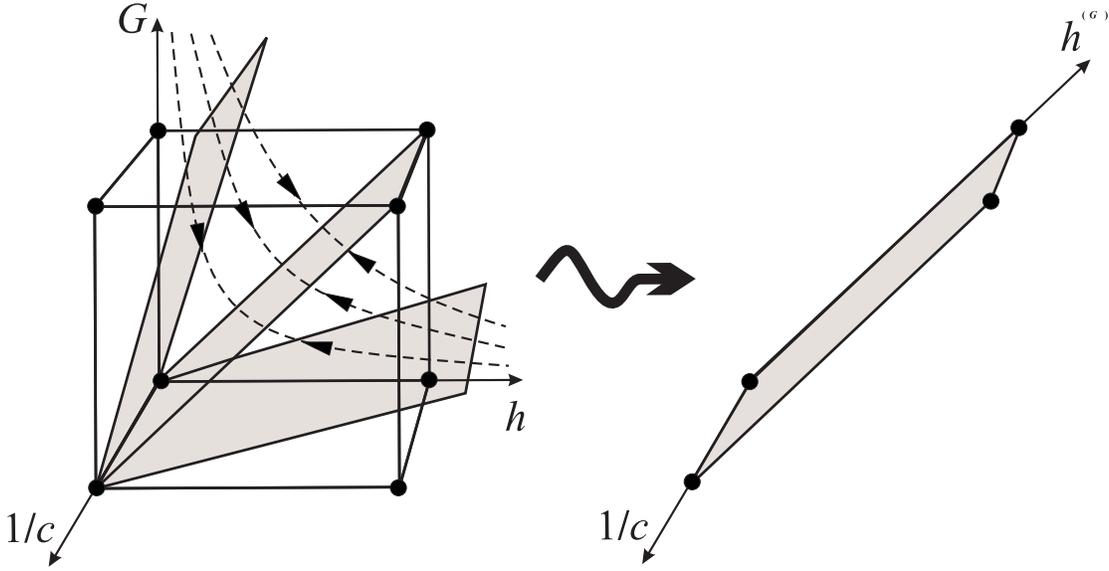,width=0.42\textwidth,angle=90}}
\caption{The cube of theories is shown in the left-hand side.
The origin $(0,0,0)$ corresponds to non-relativistic mechanics,
$(c^{-1},0,0)$ to Special Relativity,
$(0,h,0)$ to non-relativistic Quantum Mechanics,
$(0,0,G)$ to Newtonian gravity,
$(c^{-1},0,G)$ to General Relativity,
$(0,h,G)$ to ``non-relativistic Quantum Gravity",
$(c^{-1},h,0)$ to usual Quantum Field Theory, and
$(c^{-1},h,G)$ to ``Quantum Gravity''.
After using the $G$ or $h$~protocols we end up with $h^{(G)} (= G^{(h)})$
rather than with $G$ and $h$. This leads the cube of theories to be
squashed into a ``plane of theories" as shown in the right-hand side of
the figure. In the light of the $G$ protocol, for instance,
the origin $(0,0)$ corresponds to non-relativistic
mechanics, $(c^{-1},0)$ to Relativity,
$(0,h^{(G)})$ to (non-relativistic) Quantum Mechanics and
$(c^{-1},h^{(G)})$ to ``relativistic Quantum Mechanics''
(which includes Quantum Field Theory and Quantum Gravity).}
\label{squashing}
\end{figure}

\newpage
\begin{figure}[h]
\begin{center}
\mbox{\epsfig{file=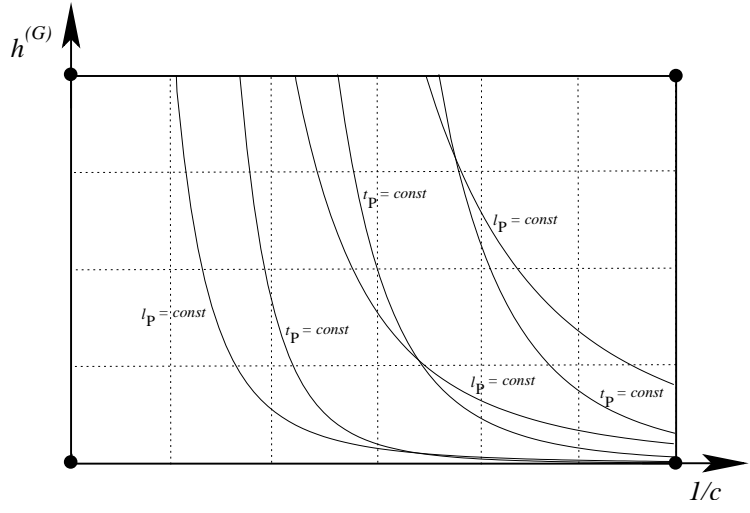,width=0.55\textwidth,angle=0}}
\end{center}
\caption{The plane of theories can be covered
by the lines of constant $h^{(G)}$ and $c^{-1}$ (horizontal and vertical dashed
lines, respectively) or, equivalently, by the lines of constant $\ell_{P}$
and $t_{P}$.}
\label{TPLP}
\end{figure}
\end{widetext}

\end{document}